# Optical polarizability of erbium-oxygen complexes in sol-gel-based silica films

Sufian Abedrabbo[1,2], Bashar Lahlouh,[2] Anthony T. Fiory,[3] and Nuggehalli M. Ravindra[4]

[1] Department of Physics, Khalifa University of Science and Technology, Abu Dhabi 51133, United Arab Emirates
[2] Department of Physics, University of Jordan, Amman 11942, Jordan
[3] Bell Labs Retired, Summit, New Jersey 07901, United States of America
[4] Department of Physics, New Jersey Institute of Technology, Newark, New Jersey 07012, United States of America
E-mail:  sufian.abedrabbo@ku.ac.ae

**Abstract**

For erbium-doped amorphous oxides, such as those that are used in compact lightwave devices interfaced with silicon, values of the refractive indices are commonly obtained empirically. This work, combining experimental and theoretical studies, examines silica as the exemplary host and the influence of erbium doping on the refractive index. Analysis of data is presented for the spectral refractive index in the ultraviolet to near infrared wavelength range of heavily erbium-doped silica thin films prepared by spin coating a sol-gel precursor on silicon and subsequent vacuum annealing. Effective medium Lorentz-Lorenz data analysis determines that the dopant component has a refractive index of 1.76 ± 0.24 with wavelength dispersion constrained to within 2 %. Considering the dopant as a localized $ErO_6$ impurity complex, a corresponding theoretical refractive index of 1.662 is derived by calculating the optical polarizability and volume of the impurity. Data presented for room-temperature (~293 K) photoluminescence in the vicinity of 1.54 µm are shown to be consistent with random variability in impurity sites. Inherent advantages of studying colloid-based materials are discussed. To the best of the authors' knowledge, such a detailed study of the refractive index associated with erbium impurities in silica is being reported for the first time in the literature.

## 1. Introduction

Erbium-containing oxides are employed in numerous applications, especially as optically active media in compact lightwave amplifiers [1] and lasers [2] fabricated and interfaced with silicon, in which atomic layer deposition of erbium-doped aluminum oxide offers a promising method for dispersing optically active impurities at high erbium concentrations [1]. Even though erbium doping is generally expected to change the refractive index of oxides through modified optical polarizabilities, scientific guidance from existing studies remains notably sparse. It is therefore reasonable that a measured refractive index was used in modeling optical device performance in [1]. In view of the variable synthesis dependence evident in deposited films, notably for aluminum oxide [3-6], this work examines the refractive index associated with erbium impurities in the more extensively studied silica as the host medium. Of relevant importance is the spectral dispersion or wavelength dependence of refractive indices.

Relatively low Er concentrations (e.g. ~400 ppm) are used in the active core of optical fiber amplifiers [7,8] and lasers [9], while substantially higher concentrations (~1% Er or more) in short waveguide structures [10] are produced by reduced-temperature methods such as sol-gel deposition of Er-doped silica [11-13] and multi-component silica-related oxides [14-17]. Erbium-doped silica-based films, suitable for lightwave devices [10,18], commonly contain one or more other cation types, including Al [14,16,19,20], Ge [20,21], Ti [10,14], Yb [14,16], Y [22], and Zr [17]. Influence of erbium doping on the refractive index is observed in some silica-based materials, e.g. as increased indices in sol-gel deposited silica-titania films [15], nominally pure silica films [12,13], and soda-lime-silicate glass [23], and decreased indices in borosilicate glass [24]. Effects of low Er concentrations on the refractive indices of glass fiber materials are generally small enough to justify their being ignored [25] or to presume the host refractive index [26] in lightwave device design. Certain doping species (e.g. Ge, Al, B, P, N, Cl, or F) are introduced to modify the refractive index and spectral dispersion in $SiO_2$ for optimal lightwave propagation [27]. The focus herein is on the refractive index in the ultraviolet to near infrared wavelength region for erbium-doped silica without co-dopants.

At erbium doping levels suitable for compact lightwave structures, measured increases observed in the refractive index of silica films [12,13] are ostensibly related to the larger optical polarizability of $Er^{3+}$ relative to $Si^{4+}$, as inferred from studies of crystalline compounds [28-30]. Thermodynamically metastable high concentrations of erbium are produced in sol-gel-based silica films by controlling the kinetics of thermal annealing [12,13], thereby achieving the primary objective of evenly dispersed $Er^{3+}$ impurities with optical activity for 1.5-µm devices [12,13,19]. The optically active species are modeled herein as $ErO_6$ impurity complexes, reflecting the 6-fold nearest-neighbor Er-O coordination in silica glass [31]. A novel calculation of the optical polarizability of $ErO_6$ species is derived in Section 2 in the context of effective medium theory of the refractive index. Section 3 describes the experimental data on refractive indices and photoluminescence. Results from the analysis of spectral refractive index and supporting photoluminescence data for erbium-doped sol-gel films and comparison with the model are presented in Section 4. Absence of wavelength dispersion in the polarizability associated with the erbium impurities, predicted for $ErO_6$ complexes and in agreement with experiment to within reasonable uncertainty values, distinguishes erbium impurities from glass-network constituents, such as germanium. To the best of the authors' knowledge, these are the first of such results reported in literature, which are discussed with prognosis for further study in Section 5.

## 2. Modeling optical refractive index

Applying effective medium theory, the optical refractive index $n$ of erbium-doped silica is theoretically determined from the Lorentz-Lorenz equation [32, 33] as

$$\frac{n^2 - 1}{n^2 + 2} = f_1 \frac{n_1^2 - 1}{n_1^2 + 2} + f_2 \frac{n_2^2 - 1}{n_2^2 + 2} \quad , \quad (1)$$

where $n_1$ is the refractive index of fused $SiO_2$ [34, 35] and $f_1$ is its molar fraction; $f_2$ and $n_2$ correspond respectively to the dopant component. Generally, the specific form of $n_2$ depends on the nature of the dopant. For example, modeling $n$ for germanium-doped silica according to equation (1) with $n_2$ as the refractive index of fused $GeO_2$ [36] and $f_1 \equiv 1 - f_2$ gives good quantitative agreement in both magnitude and wavelength dependence with the data reported in [27] (Figure S1, Supplementary data). However, in modeling $n$ for erbium-doped sol-gel film data in [13], $n_2$ turns out to be approximately constant, which is quite different from the wavelength dispersion observed for erbium oxide (see Section 4). Therefore, $n_2$ in this case originates from a fixed optical polarizability $\alpha_2$ and which, from the Lorentz-Lorenz equation, is expressed as

$$n_2 = \left(\frac{1 + 2g_2}{1 - g_2}\right)^{1/2} \quad , \quad (2)$$

where for cubic symmetry, $g_2 \equiv 4\pi\alpha_2/3V_m$ and $V_m$ is the specific molar volume (normalized to Avogadro's constant) of the dopant species. Observation of fixed or static $n_2$ is evidence for the electronic nature of $\alpha_2$, motivating its theoretical derivation for $ErO_6$ as the dopant species in the following subsection 2.1.

Empirical optical polarizabilities in dielectrics can be obtained by representing experimental $n(\lambda)$ as a function of wavelength $\lambda$ within the ultraviolet to near infrared region by the Sellmeier model with only one term [28,30],

$$n(\lambda) = \left(1 + \frac{A_0}{1 - (\lambda_0/\lambda)^2}\right)^{1/2} . \quad (3)$$

The Sellmeier parameters have physical significance through the relations $A_0 = E_d/E_0$ and $\lambda_0 = hc/E_0$, where the oscillator energy $E_0$ and the dispersion energy $E_d$ are connected to inter-band transitions and to the optical band gap [37]. The long-wavelength limit of the Sellmeier function, $n_\infty \equiv n(\lambda \to \infty) = (1+A_0)^{1/2}$, is used to define the optical polarizability $\alpha_e$ of a dielectric material in terms of the specific molar volume of the molecular formula unit [30],

$$\alpha_e = \frac{3}{4\pi} V_m \frac{n_\infty^2 - 1}{n_\infty^2 + 2} . \quad (4)$$

Applied to the host material, the refractive index of fused silica [34,35] modeled by equation (3) for 240 nm ≤ λ ≤ 840 nm (experimental range obtained optically in [13]; see Section 3), corresponds to $n_\infty \cong 1.448$. Mass density of 2.196 g/cm³ in silica corresponds to $V_m(SiO_2) = 45.43$ Å³ [28,30]. From equation (4), one has $\alpha_e(SiO_2) = 2.903$ Å³ for fused silica, a value which is utilized in Section 2.1.

### 2.1. Modeling Er in SiO₂

Information on the local structural and electronic environment of Er in glassy SiO₂ was determined for two Er concentrations (0.3 and 1.6 at.%) by Fourier transform analysis of extended X-ray absorption fine structure (EXAFS) [31]. The Er-O coordination number in SiO₂, found to be 6.0, is the same as measured by EXAFS for crystalline Er₂O₃. The Er-O bond length of 2.28 Å and root-mean-square relative displacement of 0.09Å bear close correspondence to the mean and difference of the two Er-O distances of 2.229 Å and 2.317 Å measured for Er₂O₃ [31]. Few studies of samples prepared by sol-gel methods include a determination of local Er-O structures [11,22]. In sol-gel silica powders doped with 5 mole % erbium and ≥ 10 % mole yttrium, EXAFS analysis finds Er₂O₃-like local structure and Er-O bond lengths for annealing at 900 °C [22]. It was also found from EXAFS that annealing sol-gel films at 500 °C is evidently too low a temperature for forming these Er-O bonds [11]. In the absence of EXAFS data for the Er-doped sol-gel films in [13] or similarly prepared samples, the plausibility of an Er₂O₃-like impurity for annealing at 700 °C is determined indirectly from comparative analysis of the photoluminescence spectra (~1.54 µm) for the film in reference to Er-doped bulk silica glass (Section 4.2).

Results from EXAFS for the local Er-O environment in the more lightly doped optical fiber materials are strikingly different. In SiO₂ doped solely with 0.073 – 0.075 mol. % Er₂O₃, the Er-O coordination number is 3.75 and bond length is 2.05 Å [38], while for 200 ppm Er, an exceptionally low 1.2-coordination with similarly short 2.03-Å bond is measured [20]. Both cited studies report increases in coordination numbers (to 6 or more) and bond lengths (2.32 – 2.36 Å) for SiO₂ co-doped with Al [38] or Al and Ge [20]. Luminescence in standard alumina-doped silica glass fibers also indicates varying $Er^{3+}$ sites at low erbium concentrations [39]. Thus, apart from anomalies at very low concentrations, which may indicate Er tied to low-level glass defects (evidently suppressed by the Al co-dopant), the local environment around $Er^{3+}$ impurities in $SiO_2$ otherwise bears a strong resemblance to that of $Er^{3+}$ in $Er_2O_3$.

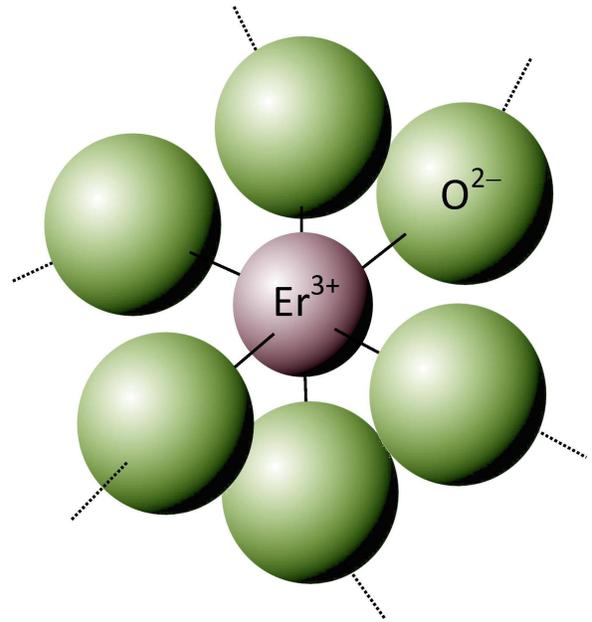

**Figure 1.** Sketch of ErO₆ impurity complex. Central $Er^{3+}$ is bonded to six $O^{2-}$ nearest neighbors (solid lines). Dotted lines represent oxygen bonding to further neighbors.

The local erbium-oxygen structure is therefore modeled as an $ErO_6$ complex, illustrated schematically in figure 1 as a distorted Er-O octahedral coordination. The six nearest-neighbor oxygens acquire electronic charge $-2e$ from the central $Er^{3+}$ and mostly $Si^{4+}$ (not illustrated) in the surrounding glassy network. Found to be essentially constant, $n_2$ evidently reflects the static molar optical polarizability $\alpha_e(ErO_6)$ and molar volume $V_m(ErO_6)$ of the $ErO_6$ complex and is expressed theoretically from equation (2) as $n_2(ErO_6) = [(1 + 2g_2)/(1 - g_2)]^{1/2}$, where $g_2 = 4\pi\alpha_e(ErO_6)/3V_m(ErO_6)$. The molar optical polarizability is calculated according to the polarizability additivity rule [29] as

$$\alpha_e(ErO_6) = \alpha_e(Er^{3+}) + (1/2)6\langle\alpha_e(O^{2-})\rangle \ . \qquad (5)$$

The factor $\langle\alpha_e(O^{2-})\rangle$ is the averaged optical polarizabilities of $O^{2-}$ in $SiO_2$ and $Er_2O_3$, denoted as $\alpha_e(O^{2-}:SiO_2)$ and $\alpha_e(O^{2-}:Er_2O_3)$, respectively, and evaluated as $\langle\alpha_e(O^{2-})\rangle = [\alpha_e(O^{2-}:SiO_2) + \alpha_e(O^{2-}:Er_2O_3)]/2$. The factor $(1/2)$ in equation (5) assumes equal sharing of $\langle\alpha_e(O^{2-})\rangle$ between the impurity complex and the surrounding host medium. From the empirical $\alpha_e(Er^{3+}) = 2.180$ Å$^3$ for 6-fold coordinated $Er^{3+}$ in [29] and $\alpha_e(Er_2O_3) = 8.293$ Å$^3$ for $Er_2O_3$ in [40], application of the additivity rule gives $\alpha_e(O^{2-}:Er_2O_3) = [\alpha_e(Er_2O_3) - 2\alpha_e(Er^{3+})]/3 = 1.311$ Å$^3$. Similarly, from $\alpha_e(Si^{4+}) = 0.333$ Å$^3$ in [29] and $\alpha_e(SiO_2) = 2.903$ Å$^3$ for fused silica, one obtains $\alpha_e(O^{2-}:SiO_2) = [\alpha_e(SiO_2) - \alpha_e(Si^{4+})]/2 = 1.285$ Å$^3$ (for comparison, $O^{2-}$ polarizabilities are 1.213 Å$^3$ and 1.250 Å$^3$ for $\alpha$- and $\beta$-quartz, respectively [29]). The 2% difference between $\alpha_e(O^{2-}:SiO_2)$ and $\alpha_e(O^{2-}:Er_2O_3)$ accounts for the small effect of $Er^{+3}$ on the optical basicity of the oxygen anions [41]. With average value $\langle\alpha_e(O^{2-})\rangle = 1.298$ Å$^3$, one obtains from equation (5), $\alpha_e(ErO_6) = 6.074$ Å$^3$. The molar volume $V_m(ErO_6)$, taking oxygen sharing into account, is estimated by subtracting the ionic volume 4.58Å$^3$ of an assumed spherical $Er^{3+}$ ion from the molar volume $V_m(Er_2O_3) = 73.33$ Å$^3$ for $Ia\bar{3}$ $Er_2O_3$ [28], yielding $V_m(ErO_6) = 68.75$ Å$^3$. The theoretical result is $g_2 = 0.370$ and application of equation (2) yields the calculated $n_2(ErO_6) = 1.662$.

## 3. Experimental section

Relevant experimental data on the optical properties of erbium-doped sol-gel silica films, along with erbium oxide and bulk erbia-silica glass as reference materials, are described in the following subsections.

### 3.1. Erbium-doped sol-gel silica films

Erbium doping levels of about 1 % are evidently sufficient for accurately resolving the $n_2$ component in $n(\lambda)$ (Section 4, Supplementary data) and readily produced in sol-gel films at temperatures substantially below glass fusion points [8]. After low-temperature curing ($\lesssim 100$ °C) [12, 13], refractive indices remain significantly smaller than those for undoped silica, owing mainly to the porous structure of gels [12] (a demonstration with $f_1 < 1$ and $n_2 = 1$ in equation (1) is shown in Supplementary data, figure S2). Thermal annealing at higher temperatures densifies the sol-gel material, increases the refractive index, and activates optical emission from $Er^{3+}$ [13]. Since N and Cl are among the dopant species known to increase the refractive index of silica [27], the sol-gel formulation based on erbium oxide and acetic acid, introduced in [13], is perhaps desirable for analysis compared to alternatives containing nitrate and chloride ions, although the colloids formed by hydrolysis of $SiC_8H_{20}O_4$ (TEOS) could contain residual C. A 700 °C anneal optimally activates $Er^{3+}$ in Er-Ge co-doped silica [21] and is used for optically active $Er^{+3}$ in sol-gel germania-silica films [42], while 670 °C activates sol-gel silica-titania films [15]. Annealing at temperatures much above 700 °C, however, may cause short-range redistribution and clustering of the Er impurities, although higher temperatures are used for depleting the trace OH impurities capable of degrading performance at 1.54 μm [12, 13] and for bringing residual porosity to very low levels [15]. Erbium-doped sol-gel films deposited on silicon and annealed at 700 °C are also optimal for enhancing band-gap radiation from the silicon substrate [43]. The authors note that the corresponding spectral refractive index data in [13] are evidently the only available such results in erbium-doped silica sufficiently complete for study according to the theoretical development of Section 2. Accordingly, analysis of these data is presented in Section 4. Experimental methods are briefly summarized below and described in further detail in the Appendix.

### 3.1.1. Sample Fabrication

Erbium in aqueous solution was prepared by dissolving 0.5 g erbium oxide powder in 4 ml acetic acid diluted with 1.6 ml deionized water and 4 ml ethanol, and stirring for 3 h at 45 °C. Two ml TEOS was then added to form the sol, which was stirred for 10 min at 80 °C. The sol was then passed through a 0.45-μm syringe filter and applied to the silicon substrate (a 25-mm square piece cut from single-side polished Czochralski p-type (100) wafer and cleaned by an RCA method); the substrate was held on a spin coater rotating at 1200 rpm for 30 s. The coated sample was cured in air for 30 min at 80 °C, after which measurements of the thickness (167.22 nm) and refractive index (1.434 at 632 nm) were taken. This method is based on generating colloidal silica from TEOS by water hydrolysis, liberating ethanol, and forming solid silica by thermal dissociation, liberating diethyl ether. The sample was then annealed in a furnace at 2-Pa vacuum and 700 °C for 1 h, followed by optical characterization measurements.

### 3.1.2. Sample Characterization

Spectral refractive index measurements were made using a Filmtek model 3000 optical metrology system [44]. The refractive index and extinction coefficient, calculated by the system over the wavelength range 240 – 830 nm in the available and used Filmtek instrument version, are determined from normal reflectivity measurements by modeling the complex permittivity of the film as damped Lorentz oscillators together using a materials database for the silicon substrate (details in Appendix). Reflectance of the film annealed at 700 °C, emulated to a root-mean-square accuracy of $2.85 \times 10^{-3}$, determines its thickness as $d = 126.12$ nm and refractive index $n = 1.471$ and extinction coefficient $k < 10^{-4}$ at wavelength $\lambda = 632$ nm. Data for other films, which were annealed at 550, 600, and 650 °C, are $d = 141$, 138, and 138 nm, and $n = 1.441$, 1.445, and 1.446 at $\lambda = 632$ nm, respectively.

Photoluminescence measurements were carried out at room temperature (~293 K) using a Horiba Jobin-Yvon model Fluorolog-3 spectrophotometer and a liquid-nitrogen cooled InGaAs photodetector, sensitive to wavelengths up to about 1600 nm. The spectrum is taken with photoexcitation at 532 nm, corresponding to optimal excitation of the $^4S_{3/2}$-$^4H_{11/2}$ manifold of $Er^{3+}$ (excitation matrix data in Supplementary data, figure S7). Spectral signals were corrected by subtracting the dark current baseline and dividing by InGaAs responsivity (the raw data are shown in [13] and figure S7). Reference photoluminescence spectra at room temperature were measured with a photomultiplier detector for a commercial erbia-silica glass (Er-Si-O glass) containing 20 ± 2 mol. % erbium oxide and proprietary concentrations (1 – 10 wgt. %) of the oxides of B, Ba, K, Li, Na, and Zn [45]. This Er-Si-O glass is selected because it emulates the important materials attributes of (1) high Er concentration, (2) silica-based host, and (3) optically active $Er^{+3}$ centers.

### 3.2. Erbium oxide

The refractive index of erbium oxide, a possible candidate for $n_2$ in equation (1), is modeled parametrically by equation (3) from empirical data. For crystalline $Er_2O_3$, $n_\infty = 1.923$ is derived in [40] from linear interpolation of prism refraction data for other rare-earth oxides ($\lambda_0 = 138$ nm by similar interpolation) and $n_\infty = 1.888$, 1.930 and 1.95 from additional sources are listed in [28]. The average and ± standard deviation value is $n_\infty = 1.923 \pm 0.026$. For thin film forms of $Er_2O_3$, reported results for $n(\lambda)$ obtain $A_0 = 2.94$ and $\lambda_0 = 138$ nm (amorphous film [46]), $A_0 = 2.49$ and $\lambda_0 = 128$ nm (polycrystalline film [47]), $A_0 = 2.65$ and $\lambda_0 = 150$ nm (amorphous film [48]), and $A_0 = 2.32$ and $\lambda_0 \approx 128$ nm estimated here from $n(\lambda)$ curves for Er films oxidized at 700 and 800 °C [49]. Averaging the film data for $A_0$ yields the result $n_\infty = 1.896 \pm 0.069$. Parameters obtained by averaging the crystal and film results are $A_0 = 2.65$

± 0.14, $\lambda_0$ = 136 ± 1 nm and $n_\infty$ = 1.909 ± 0.036, and are used to model the refractive index of erbium oxide.

## 4. Results and analysis

The following sections present results on the analysis of the data for refractive index and photoluminescence, and the theoretical refractive index of erbium-doped silica extrapolated into the infrared.

*4.1. Refractive index*

Data for the spectral refractive index $n(\lambda)$ of the SiO$_2$:Er sol-gel film (126-nm thickness, 700 °C anneal) are shown in figure 2 as circle symbols. The lower solid curve shows the index of fused SiO$_2$. In order to model the film data according to equation (1), the one-term Sellmeier function of equation (3) is used to represent the possible dispersion in $n_2$ vs. $\lambda$ for the dopant component, denoted as ErO$_x$. Least squares minimization with parameters $A$, $\lambda_0$, $f_1$, and $f_2 \equiv (1-f_1)$ finds near zero dispersion parameter, $\lambda_0 = (-2 \pm 6)$ nm, revealing that $n_2$ of ErO$_x$ is indistinguishable from a constant function of wavelength. Consequently, $n_2$ is treated as a single parameter in equation (1), equivalent to fixing $\lambda_0 \equiv 0$. Fitting the data for $n(\lambda)$ with two independent parameters yields $f_1$ = 0.947 ± 0.004 and $n_2$ = 1.76 ± 0.24. By identity, the dopant fraction is $f_2 \equiv 1- f_1 = 0.053 \pm 0.004$. The fitted function, shown as the curve underneath the data points in figure 2, deviates from the data with a root-mean-square (rms) average of 0.0004.

Deviation between fitted function and the data are examined by holding $f_1$ constant at 0.947 and solving equation (1) for $n_2$ at individual wavelengths. This procedure extracts the wavelength variation in $n_2$. The result is shown as the dotted curve in figure 3, where the horizontal solid line represents the fitted constant for $n_2$. The small non-monotonic wavelength variations in $n_2$, constrained to a 2 % range, are suggestive of systematic uncertainty. Values of $n_2$ vary with a standard deviation of 0.0096, corresponding to 4 % of the uncertainty in the fitted constant for $n_2$. The dashed curve in figure 3 is the Sellmeier model function representing Er$_2$O$_3$ ($A_0$ = 2.65, $\lambda_0$ = 136 nm), shown for comparison, where the refractive index monotonically increases by up to 19 % at short wavelengths.

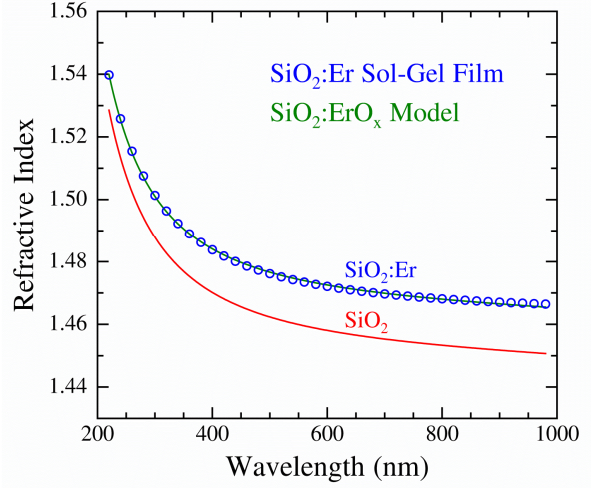

**Figure 2.** Refractive index versus vacuum wavelength measured for 126-nm SiO$_2$:Er sol-gel film on silicon annealed at 700 °C (circles), model function (solid curve through circles), and for bulk SiO$_2$ (lower solid curve).

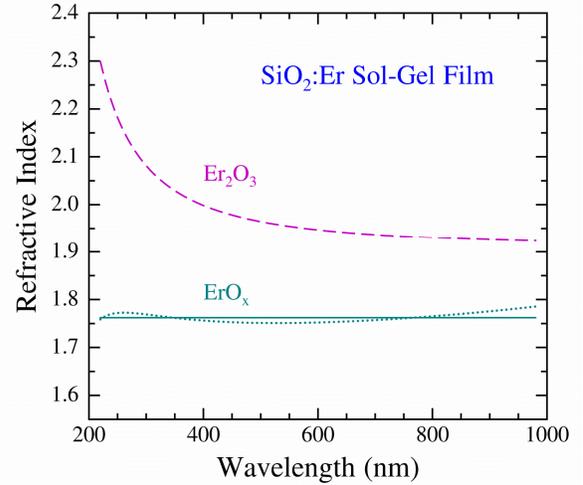

**Figure 3.** Refractive index of ErO$_x$ component (dotted line) and mean value of 1.76 (horizontal line) from fitting the refractive index of SiO$_2$:Er sol-gel film. Refractive index of pure Er$_2$O$_3$ is shown by dashed curve.

Possible effects of residual film porosity, observable in some sol-gel materials [15,50], are examined by introducing the inequality $f_2 < (1-f_1)$ in equation (1), which is equivalent to adding a zero-valued third term to the right side with fraction $f_3 = 1- f_1- f_2$ and refractive index $n_3$ = 1.

In order to model uncertainties associated with $f_3$, the maximum $n_2$ is presumed to be limited by $n_\infty = 1.909$ for $Er_2O_3$, which lies within the aforestated uncertainty in $n_2$ and consequently limits $f_3$ to 0.0064 or less. The rms deviation in $n_2$ from its mean remains less than 0.01 over this range of uncertainty in $f_3$, verifying that the very weak variation in $n_2$ with wavelength is a stable property of the data. This finding is reproduced by treating $f_2$ as an independent parameter, yielding $f_1 \approx 0.95$, $f_2 \approx 0.04$, $f_3 = 1-f_1-f_2 \approx 0.01$, and $n_2 \approx 1.85$. Thus, the upturn in the index at short wavelength exhibited in pure $Er_2O_3$ (dashed curve in figure 3) is not in evidence for $n_2$. Expressing this distinction statistically, $n(\lambda)$ for $Er_2O_3$ deviates by the much larger 0.08 rms from its mean.

Similar analysis was applied to the refractive index data in [13] for the film annealed at 650 °C, for which the refractive index is smaller and the thickness is larger. Film porosity effects are readily quantified by treating $f_2$ as an independent parameter, along with $f_1$ and $n_2$. The results are $f_1 \approx 0.87$, $f_2 \approx 0.09$, $f_3 = 1-f_1-f_2 \approx 0.04$, and $n_2 \approx 1.76$. Owing to parameter correlations in the model function, uncertainties are undetermined when treating the spectral dependence of $n$ with three independent parameters. However, consistent results are obtained from the spectral refractive index obtained for annealing at 650 °C and 700 °C.

*4.2. Photoluminescence*

Particular evidence concerning the local structure of the $ErO_x$ impurity, in view of the $ErO_6$ structure shown in figure 1, is available from the $Er^{+3}$ photoluminescence near 1.54 μm. Room-temperature (~293 K) photoluminescence from the erbium-doped sol-gel film annealed at 700 °C is presented in figure 4 (circle symbols), showing that the wavelength dependence exhibits a single smooth maximum. Observation of this spectral shape is virtually independent of anneal temperatures up to at least 850 °C [13], verifying that the photoluminescence originates from $Er^{+3}$ in dynamically stable structures. This corroborates the low temperature spectroscopy classic study by C.C. Robinson [51] where it is concluded that in undoped silicate glasses the sixfold Er-O coordination dominates. The allowed Stark split

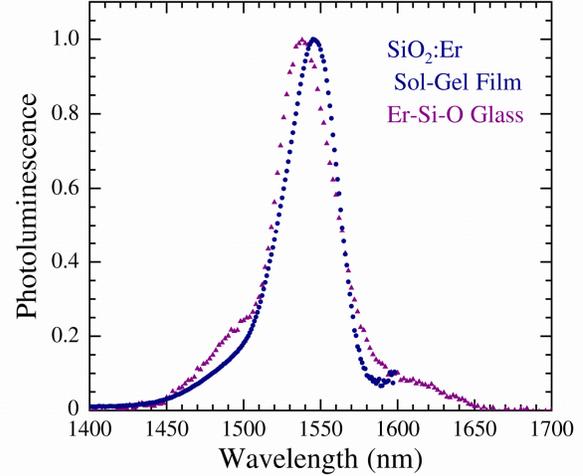

**Figure 4.** Room-temperature photoluminescence spectra of erbium-doped sol-gel silica film (filled circles) and bulk erbia-silica glass (filled triangles), normalized to unity maxima.

$^4I_{13/2} \to {^4I_{15/2}}$ emission transitions from $Er^{+3}$ in silica are typically manifested as two dominant maxima near 1535 and 1550 nm [8], which are clearly resolved at low Er concentrations in fused silica [8,25] and for the Er-doped $SiO_2$ samples studied by EXAFS in [31]. The film spectrum has an inhomogeneous broadening of 14 ± 1 nm that merges the emissions near 1535 and 1550 nm such that a single maximum is observed. A similar broadening effect occurs in the photoluminescence from bulk Er-Si-O glass with high Er concentration (~10 mol. % as $ErO_x$), which is included for comparison in figure 4 (triangle symbols). The spectra from both the film and the samples of Er-doped $SiO_2$ reported in [25,31] show a smooth variation from ~1455 nm to ~1500 nm corresponding to thermal occupation of higher Stark levels of the $I_{13/2}$ manifold [52]. This emission appears as a pronounced shoulder in the bulk glass sample similarly to silicate glass [8,52]. At 5.3 % Er concentration, there is a statistical probability of 0.28 that an oxygen in an $ErO_6$ structure bridges another Er. Theoretically, the effect on $n_2$ is very small, owing to the 2% differences in $O^{2-}$ polarizabilities. However, inhomogeneous broadening of $Er^{+3}$ emissions increases with Er concentration, as shown by studies of sol-gel material with Er concentrations above 1 % [19]. Comparisons of photoluminescence spectra in the aforecited [19,25,31] to

figure 4 are illustrated in Supplementary data figure S3. Analysis of the film photoluminescence is shown in Supplementary data figure S4.

*4.3. Infrared refractive index*

Theoretical curves are presented in figure 5, which extend into the infrared region beyond the range of available data, showing dependences on vacuum wavelength $\lambda$ of the refractive index $n$ as solid curves and the group index $n_g = n - \lambda dn/d\lambda$ as dotted curves. The upper two green curves are calculated for $SiO_2$:Er with atomic concentration $c$ = 0.05 and the lower two red curves are for undoped $SiO_2$. Equation (1) is used with empirical refractive index data for $SiO_2$ in [34] for $n_1$, the theoretical $n_2 = 1.662$ determined for the $ErO_6$ complex in Section 2.1, $f_1 = 1 - c$, and $f_2 = c$. The curves for $n_g$ exhibit minima corresponding to the inflection point near $\lambda_D \approx 1273$ nm in the spectral dispersion of $n_1$ for $SiO_2$ [34]. For $\lambda = 1540$ nm, within the lightwave communications band, the refractive index values are $n = 1.454$ and $n_1 = 1.444$ for Er doped and undoped $SiO_2$, respectively, constituting an increase of 0.010. The corresponding group indexes are 1.472 and 1.463, where the increase is 0.009.

Corrections for resonant optical absorption from $Er^{+3}$, although not included in the above analysis, may be estimated from the average absorption coefficient $\alpha \sim 10^2$ cm$^{-1}$ measured in the 1540 nm band for a crystalline erbium oxide film, where the spectral refractive index shows a smooth variation with wavelength [53]. Scaling by Er concentration predicts an extinction coefficient $k < 10^{-4}$ for the $SiO_2$:Er film and thus a small correction in $n$ (fourth decimal place or less). The group index is expected to increase up to $\sim 6 \times 10^{-5}$ near 1540 nm (Supplementary data figure S6). An available experimental assessment is found in the measurements of $n$ and $n_g$ for aluminum oxide films doped with 4.9% Er presented in [1], which vary smoothly in the vicinity of 1540 nm and are used therein for modal analysis of hybrid waveguides.

For Er doping of $SiO_2$, the increases in $n$ and $n_g$ are theoretically determined for small $c$ by calculating the partial derivatives $\partial n/\partial c$ and $\partial n_g/\partial c$. Normalizing to the fractional change in refractive index by defining $\delta n/cn \equiv n^{-1}\partial n/\partial c$, the result from equation (1) is

$$\frac{\delta n}{cn} = \frac{(n_1^2 + 2)^2}{6n_1^2}(g_2 - g_1) , \qquad (6)$$

where $g_i = (n_i^2+2)^{-1}(n_i^2-1)$ for $i = 1$ and 2. The fractional change in group index is similarly defined as $\delta n_g/cn_g \equiv n_g^{-1}\partial n_g/\partial c$, which, utilizing equation (6), may be written as

$$\frac{\delta n_g}{cn_g} = n_g^{-1}n\left[\frac{\delta n}{cn} - \lambda\frac{\partial}{\partial \lambda}\frac{\delta n}{cn}\right] . \qquad (7)$$

Results for the fractional changes (increases) calculated from equations (6) and (7) are shown in figure 6. While $\delta n/cn$ for the refractive index increases monotonically with wavelength, $\delta n_g/cn_g$ for the group index exhibits a maximum at $\lambda = 1271$ nm. Theoretically, these fractional increases at $\lambda = 1540$ nm are $\delta n/cn = 0.139$ and $\delta n_g/cn_g = 0.128$. At $c = 0.05$, the fractional increases change slightly, owing to non-linear dependence on $c$, to $\delta n/cn = 0.140$ and $\delta n_g/cn_g = 0.127$.

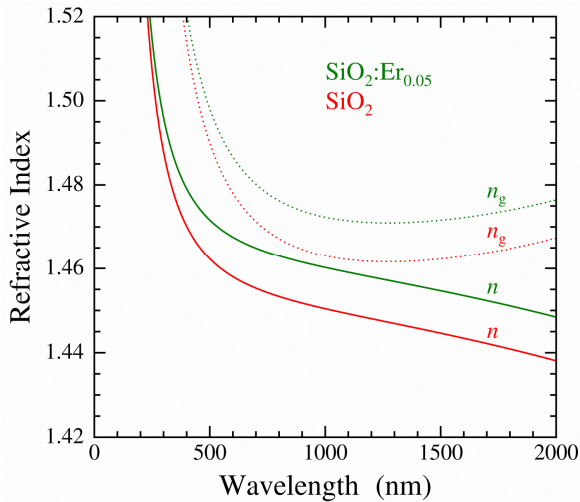

**Figure 5.** Theoretical dependence on vacuum wavelength of the refractive index $n$ (solid curves) and group index $n_g$ (dotted curves) for $SiO_2$ doped with 0.05 atomic concentration ($SiO_2$:$Er_{0.05}$, green curves) and undoped ($SiO_2$, red curves).

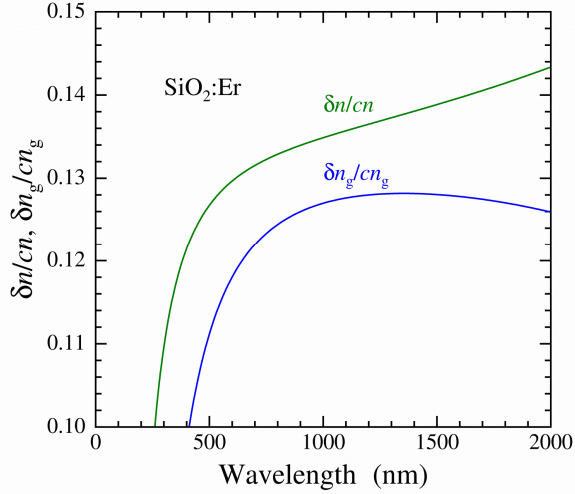

**Figure 6.** Theoretical dependence on vacuum wavelength of the fractional increase per unit concentration in the refractive index ($\delta n/cn$, upper curve) and group index ($\delta n_g/cn_g$, lower curve) for Er-doped $SiO_2$.

The wavelength of zero dispersion, corresponding to $d^2n/d\lambda^2 = 0$ and occurring at $\lambda_D \approx 1273$ nm in $SiO_2$, is an important materials metric for mode propagation in optical fibers [27]. The theoretical function for $n$ predicts that Er doping at $c = 0.05$ leaves $\lambda_D$ unchanged to within 1 nm. (Supplementary data, figure S5).

## 5. Discussion

Modeling the dopant species with static optical polarizability, as determined for $ErO_6$ by equation (5), is shown to be consistent with the near wavelength independence in the dopant component found in the analysis of refractive index data according to equation (1). The dopant concentration of $(5.3 \pm 0.4)$ % obtained from the two-parameter fit to the data is about consistent with the erbium doping level of 6 at. % stated in [13]. The theoretical magnitude of $n_2(ErO_6)$ falls within the $\pm 13.6$ % uncertainty in the fitted $n_2$ parameter.

Near absence of dispersion in $n_2$ shows that erbium-doped silica likely contains a dopant species like $ErO_6$, rather than responding like an admixture of $SiO_2$ and $Er_2O_3$ glass components, a conclusion drawn from samples annealed at 650 °C and 700 °C. Annealing at 550 °C produces nearly the same lower refractive index as for films cured at 80 °C, which is consistent with anomalous Er-O coordination and strong thermal quenching of photoluminescence reported for the 500-°C anneal in [11]. Photoluminescence data suggest that the $ErO_x$ complex is stable against Er diffusion and/or clustering for annealing temperatures up to 850 °C [13]. Thus the prediction of $n_2 = 1.662$ from Section 2.1 is expected to apply for films annealed in the 650 °C –850 °C temperature range. Whether mixed-oxide ($Er_2O_2$-$SiO_2$) effective medium behavior from Er clustering appears in the refractive index for higher annealing temperatures is undetermined although possible.

The limited availability data clearly hampers the ability to draw general inferences on the effect of erbium doping on the refractive index of oxide materials. Importantly, theory for the form of the refractive index for erbium-doped $Al_2O_3$, such as in [1], remains an open question. Although similarly prepared erbium-doped sol-gel silica films are studied in [11], there are unfortunately no reported refractive index data for annealed samples.

The following discussion considers additional information from published refractive index data for other amorphous oxide hosts doped with erbium (analysis details in Supplementary data). Most of the studies report the refractive index at a single wavelength, and therefore minimal information is available on dispersion in $n_2$. For the Na-Ca-Si-Er-oxide melt-quench glasses studied in [23], the refractive index as function of erbium content, modeled by equation (1), indicates $n_2 = 2.2 \pm 0.1$. For the Zn-B-Si-Er-oxide glasses studied in [24], the refractive index decreases with erbium content, indicating $n_2 = 2.0 \pm 0.3$. Interpreting these values of $n_2$ by extending the model of Section 2.1 involves taking into account erbium-induced changes in oxide polarizability and bonding in these ternary-oxide hosts [23,24] on a level of complexity that is beyond the scope of the current work. For the germania and germania-silica sol-gel films, including co-doping with Al and P, studied in [54], experimental resolution is insufficient to discern changes from erbium doping. Refractive index data reported for erbium-doped sol-gel silica alloyed with hafnia [55], titania [15] and germania [42], have similarly insufficient experimental resolution. Erbium doping of the Te-B-Zn-oxide glasses studied in [56] is interpreted as forming non-bridging oxygens, which obscures changes associated with cation polarizabilities in this silica-free tellurite-based glass.

Clearly, a case can be made for extended studies of erbium-doped oxides, in view of the limited information available in prior experimental and theoretical work. Points for consideration in experimental research obviously include accurate measurements of wavelength dependence of the refractive index and the extinction coefficient, as well as the parameters involved in sample preparation. Samples prepared from colloid precursors of silica and the erbium dopant are advantageous in that annealing kinetics are readily managed at reduced process temperatures [13], when compared to melt-quench glass processing. Owing to inherent uncertainties in analysis based on modeling the reflectivity of dielectric films on substrates, monolithic sol-gel samples are potential alternatives for improving the experimental resolution in the refractive index components. Other low-temperature methods for preparing heavily erbium doped films utilize ion-beam mixing in combination with co-deposition [57] or ion implantation [58]. Accurate measurements of extinction coefficient spectra are of value in determining the effects of erbium doping on absorption edges, optical band gaps, and associated modifications in the glassy framework. Extending studies to include co-doping components, such as Al and Ge, addresses materials bearing technological value.

**Conclusions**

The optical polarizability of erbium doped amorphous $SiO_2$ is proposed to include the optical polarizabilities of $ErO_6$ impurity complexes in the silica host. A novel theoretical calculation of the optical molar polarizability and volume of an $ErO_6$ complex is introduced, predicting a constant component in the refractive index with 1.662 value. In comparing the calculation to experiment, the spectral refractive index of a studied Er-doped sol-gel silica film is analyzed by the Lorentz-Lorenz effective medium model. The result is a (5.3 ± 0.4) % inclusion of a dopant component with a refractive index of magnitude 1.76 ± 0.24. Wavelength dispersion is confined to a 2.0 % range, irrespective of uncertainties in the refractive index that include sol-gel porosity and dopant concentration. The model agrees with the experimentally observed near absence of dispersion in the erbium-dopant component and the calculated magnitude is consistent within experimental uncertainty. Room temperature photoluminescence spectra observed in both Er-doped sol-gel silica and a bulk erbium-silicon oxide glass show a single broadened peak, consistent with randomly distributed $Er^{3+}$ sites among the $ErO_6$ impurity complexes.

**Acknowledgements**

This publication is based on work supported by the Khalifa University of Science and Technology under Award No. CIRA-2019-043, the Abu Dhabi Award for Research Excellence (AARE) 2018 (Project Contract No. AARE18-064), The University of Jordan, and New Jersey Institute of Technology. The authors would like to acknowledge J.C. Hensel for helpful remarks and enlightening discussion on intra-band transitions. The authors thank Horiba Jobin-Yvon, Edison, New Jersey, USA, for the measurements of photoluminescence from the Er-Si-O glass samples.

Preprint of J. Phys. D: Appl. Phys. **54** 135101 (2021) Doi: 10.1088/1361-6463/abd5e4

**Appendix**

Reflectance *vs*. wavelength data (240 – 830 nm) obtained by the Filmtek instrument for $SiO_2$:Er films are shown in figure 7. Blue curves are measured reflectance and red curves are model simulations. Reflectance for the film cured in air at 80 °C is shown in panel (a) and after vacuum furnace anneal at 700 °C in panel (b). The rms deviations between simulations and data, expressed in percentage reflectance units, are 0.238 in (a) and 0.285 in (b). Refractive index *n* (red) and extinction coefficient *k* (blue) vs. wavelength for the $SiO_2$:Er film after curing in air at 80 °C is shown in panel (c) and after furnace anneal at 700 °C in panel (d).

Data simulations, performed by internal Filmtek software [44], use a generalization of the Lorentz oscillator material model for the complex dielectric function and refractive index of the film as a function of photon energy $E$,

$$\varepsilon = \varepsilon_1 + i\varepsilon_2 = (n + ik)^2$$
$$= 1 + \sum_{j=1}^{m} \frac{A_j^2}{(E_c)_j^2 - E^2 + i(E_v)_j E},$$

where $A_j$ is the amplitude, $(E_c)_j$ and $(E_v)_j$ are denoted center and vibration energies, respectively, and $j$ is the oscillator index. A data-based model is used for the spectral refractive index and extinction coefficient for the Si substrate. Values of parameters are determined within the Filmtek software, which employs a damping coefficient in the generalization of the Lorentz oscillator model, as described in [44]. Parameters for the annealed film are thickness $d$ = 126.12 nm, roughness = 0.0 nm, $m = 1$ (software selected), with $A_1 = 14.689609$, $E_{c1} = 13.753732$ eV, $E_{v1} = 0.197516$ eV, and damping coefficient of 0.904124. Simulation of $n$ is shown in figure 2. Since Er doping makes negligible contributions to wavelength dispersion in the refractive index, the finite values found for $k$ arise almost entirely from the wavelength dispersion in the refractive index of the host silica material. The finding $k \ll n$ also permits using the form $n = (\varepsilon_1)^{1/2}$ and neglecting $k$ in theoretical modeling for the wavelength region of the data.

Infrared transmittance at wavelengths from 1000 to 1800 nm for samples of Er-doped sol–gel films on Si annealed at various temperatures are reported in [43], which are treated qualitatively therein in the absence of corresponding reflectance measurements.

Supplementary material appears after the references.

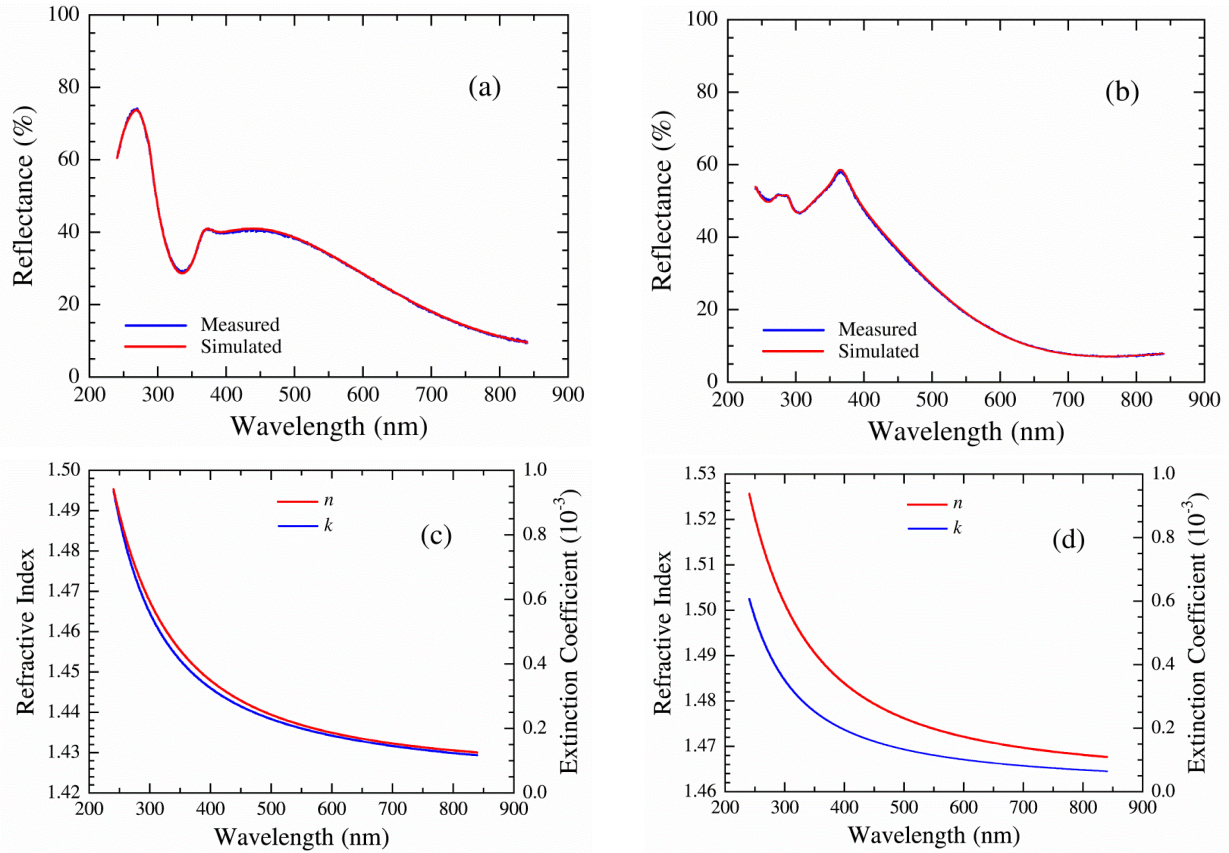

**Figure 7.** Reflectance spectra of SiO$_2$:Er films (a) after curing at 80 °C and (b) after annealing at 700°C; measurements are shown in blue, simulations are in red. Corresponding spectra for film refractive index ($n$, red, left scale) and extinction coefficient ($k$, blue, right scale) are shown in panels (c) and (d), respectively.

## Supplementary Material

Supplementary material includes Figs. S1 – S9 cited in the article and experimental data and analysis of refractive indices, reflectivity, and photoluminescence.

Refractive indices of glasses, measured using a prism technique, were reported for fiber optic SiO$_2$ materials doped with Ge, Cl and other species in [S1]. The data were presented as a table of fitted Sellmeier parameters using 3 terms. Figure S1 shows results for SiO$_2$ (red curve) and two samples doped with Ge concentrations of 4.5 mol. % (green squares) and 11.6% (blue circles). Fitted model curves are drawn through the data for the Ge-doped samples, based on equation (1) in the article using this SiO$_2$ curve for $n_1$, published data for GeO$_2$ [S2], and $f_2 = 1 - f_1 = 0.045$ (green curve) or $f_2 = 0.116$ (magenta curve). Agreement between model functions and data appears to be rather acceptable. Systematic differences suggest that the data display slightly weaker wavelength dispersion than calculated.

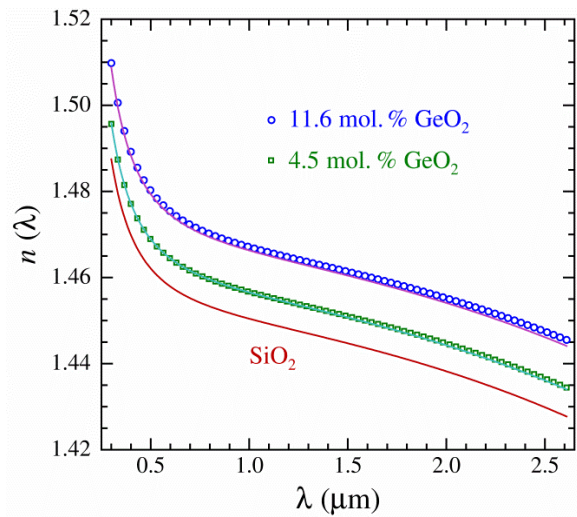

**Figure S1.** Refractive index of pure and Ge-doped SiO$_2$ glasses [S1].

Figure S2 shows the refractive index data for the erbium-doped sol-gel silica film after curing in air and before vacuum annealing (triangle symbols) and a theoretical model function (green curve) derived from equation (1) in the article. The refractive index $n_1$ of the unannealed SiO$_2$ medium is allowed to be shifted in wavelength by $\Delta\lambda$, e.g. $n_1 = n_{SiO2}(\lambda+\Delta\lambda)$, where $n_{SiO2}(\lambda)$ for fused silica is the red dashed curve [S3], and fixed inclusion index $n_2 = 1$ is used to model porosity. The results of least squares deviation minimization with two parameters are $f_2 = 0.0441 \pm 0.0002$, $\Delta\lambda = -14.1 \pm 0.6$ nm, and rms deviation 0.0003. These results suggest a reduced optical gap in the unannealed sol-gel silica, in that the one Sellmeier oscillator parameter, where $\lambda_0 = 93$ nm is obtained for $n_{SiO2}(\lambda)$, is increased to 107 nm, or about 15%. After annealing at 700 °C: $\Delta\lambda = 0$.

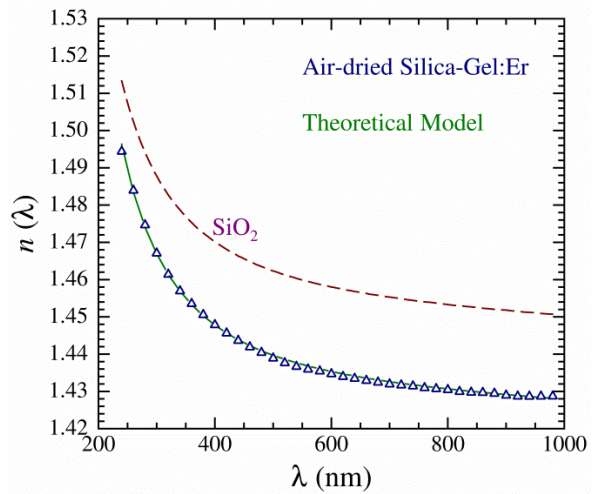

**Figure S2.** Refractive index of air-dried Er-doped silica gel film and fused SiO$_2$ glasses.

Photoluminescence data for erbium-doped silica fiber material (~200 ppm Er) has been reported in figure 4 of Wang *et al*. [S4], for fused silica glass (0.3 at. % Er) in figure 1 of Marcus *et al*. [S5], and for sol-gel silica doped with 10 wgt. % Er (1.7 mol. %, heated to 800 °C) in figure 4 of Stone *et al*. [S6]. Tracings from the figures in the references are shown in green color in figures S3 (a), (b), and (c), respectively, on scales matching figure 4 for Er:SiO$_2$ sol-gel film and Er-Si-O-glass in the article, displayed here as overlay. Wavelengths at photoluminescence maxima are 1546 nm and 1538 nm with overall spectral widths (full widths at half maximum) of 42 nm 45 nm for Silica:Er and Er-Si-O glass, respectively. For the glass fiber in (a), the maxima are at 1535 and 1552 nm with individual widths ~ 10 nm. For the fused glass in (b), the two maxima are at 1536 nm and 1550 nm, and the overall spectral width is 32 nm. For the sol-gel film in (c), the two maxima are at 1535 and 1547 nm and the overall spectral width is 52 nm.

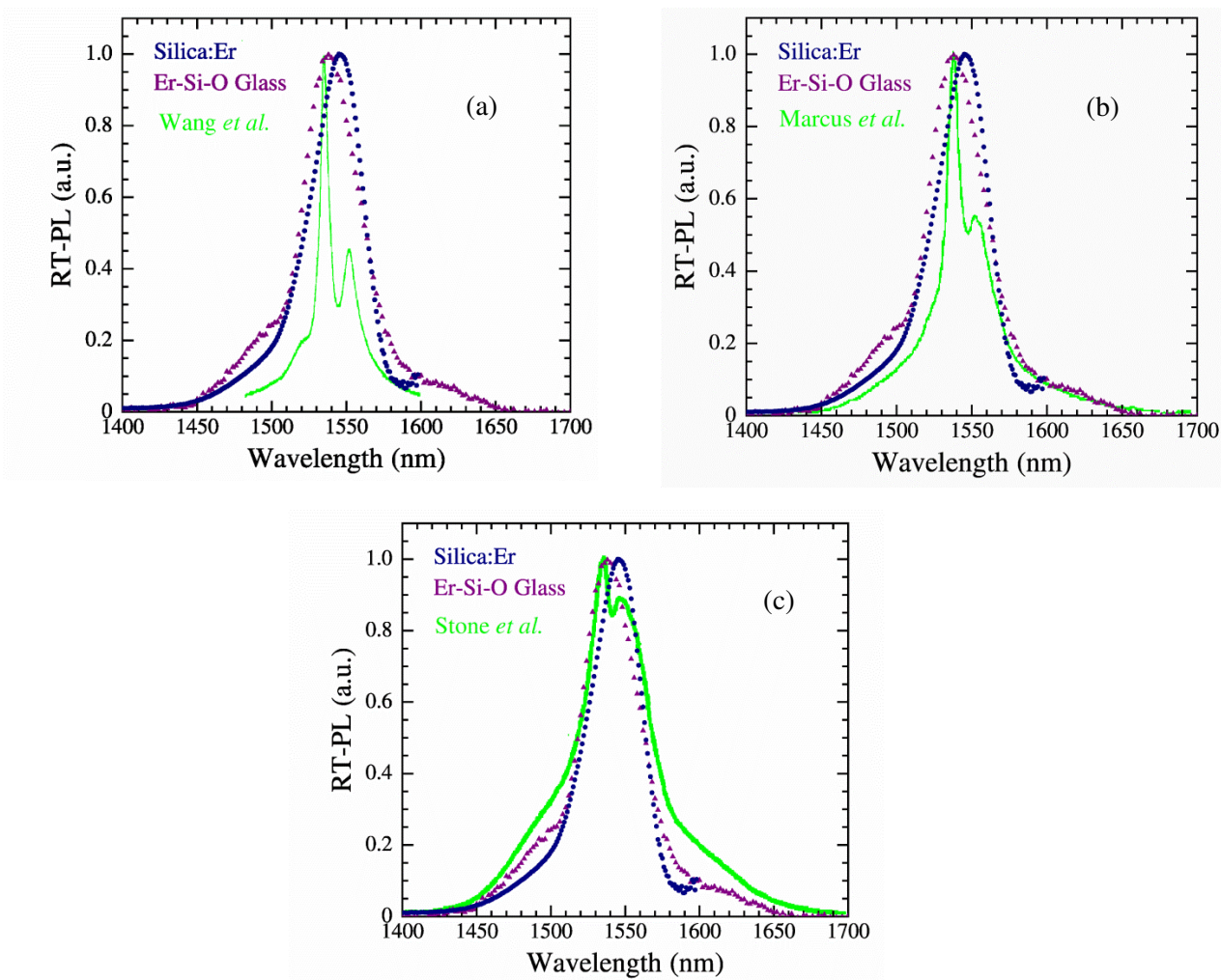

**Figure S3.** Room-temperature photoluminescence spectra of erbium-doped sol-gel silica film (filled circles) and bulk erbia-silica glass (filled triangles), from figure 4 of the article together with photoluminescence spectra of erbium-doped silica (a) fiber (Wang *et al*. [S4]), (b) EXAFS sample (Marcus *et al*. [S5]) and (c) sol-gel (Stone *et al*. [S6]), shown in green. Spectra are normalized to unity maximum.

The photoluminescence spectrum of the erbium-doped sol gel film is shown in figure S4 (symbols), and a model spectrum comprising a Lorentzian and two Gaussians (red curve). The Lorentzian lineshape has the form $[1+\Gamma^{-2}(\lambda-\lambda_1)^2]^{-1}$ with $\lambda_1 = 1505$ nm and $\Gamma = 30$ nm and represents the homogeneously broadened thermally excited intra-4$f$ $Er^{+3}$ emissions. The two Gaussians of form $\exp[-\sigma^{-2}(\lambda-\lambda_i)^2/2]$ with $\lambda_2 = 1536$ nm, $\lambda_2 = 1550$ nm, and $\sigma = 15$ nm emulate inhomogeneous broadening of the two peaks observed in the EXAFS sample (figure S3(b)).

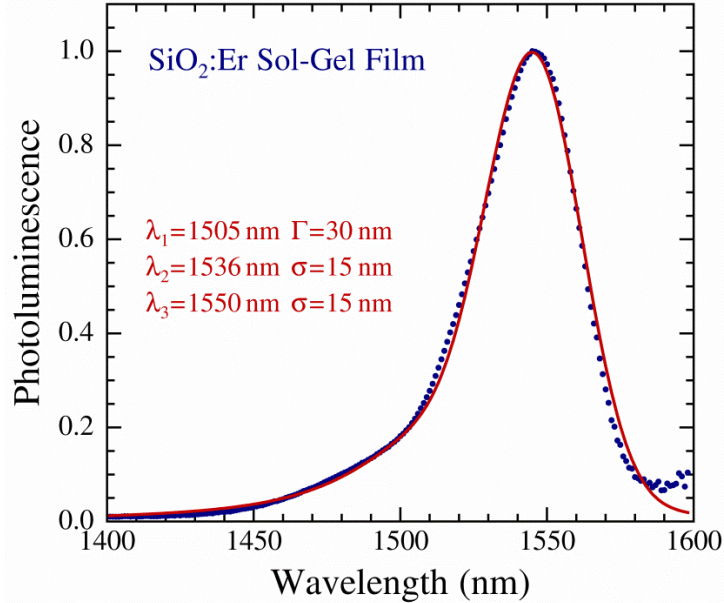

**Figure S4.** Photoluminescence at ~293 K from erbium doped sol-gel film annealed at 700 °C (circle symbols) and theory (red curve) for emissions of Lorentzian ($\lambda_1$) and Gaussian ($\lambda_2$, $\lambda_3$) lineshapes; parameters are given in legend.

Zero crossing of the dispersion coefficient for material dispersion, $D_\lambda = -\lambda d^2n/d\lambda^2$, occurs at the inflection point of $n(\lambda)$; the corresponding wavelength $\lambda_D$ is a parameter governing mode propagation in optical fibers [S1]. The $D_\lambda$ function obtained for the theoretical refractive index for erbium concentration $c = 0.05$ is shown as the blue curve in figure S5. For comparison, the $D_\lambda$ function calculated from data for fused silica [S7] is shown as the red curve. Zero crossing occurs at $\lambda_D = 1272.6$ nm for fused $SiO_2$ and theoretically at $\lambda_D = 1271.1$ nm for $SiO_2:Er_{0.05}$.

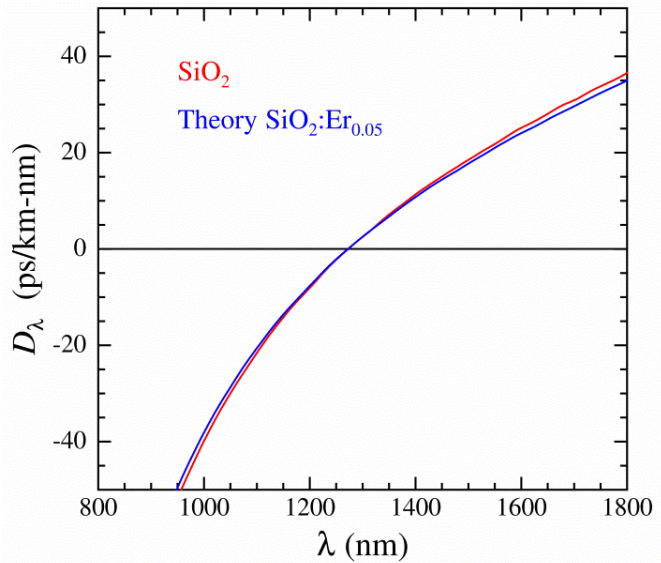

**Figure S5.** Dispersion coefficient for material dispersion $D_\lambda$ for $SiO_2$ (red curve) and $SiO_2$ doped with 5 at. % Er (theory, blue curve)

Theoretical group index for $SiO_2$ doped with 5% Er is shown on expanded scales in figure S6. The estimated correction for intra $4f$ $Er^{3+}$ absorption near 1540 nm wavelength is estimated as 5% of the resonant absorption observed in crystalline films of $Er_2O_3$ [S8].

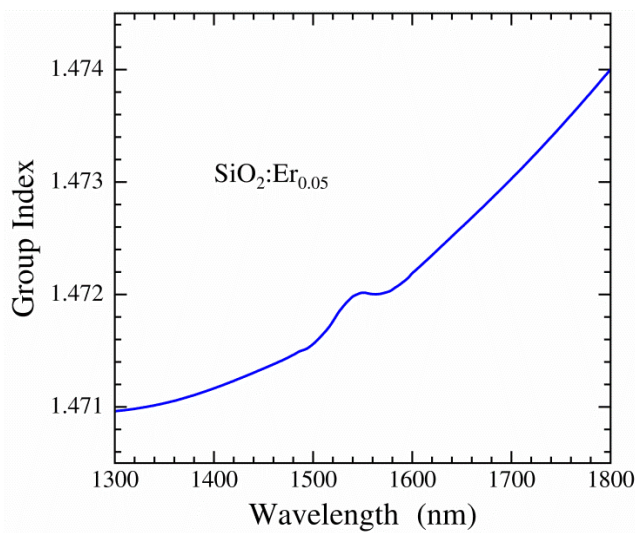

**Figure S6.** Theoretical group index of $SiO_2$ doped with 5% Er as function of wavelength.

Photoluminescence from the $SiO_2$:Er film annealed at 700 °C was measured at room temperature (~293 K) as functions of excitation and emission wavelengths. The detector signal is divided by excitation power. The ensuing excitation-emission matrix is shown in figure S7. Maximum photoluminescence signal is observed at 532 nm excitation wavelength, which is used to produce the spectra shown in figure 4 of the article and reproduced in figure S3.

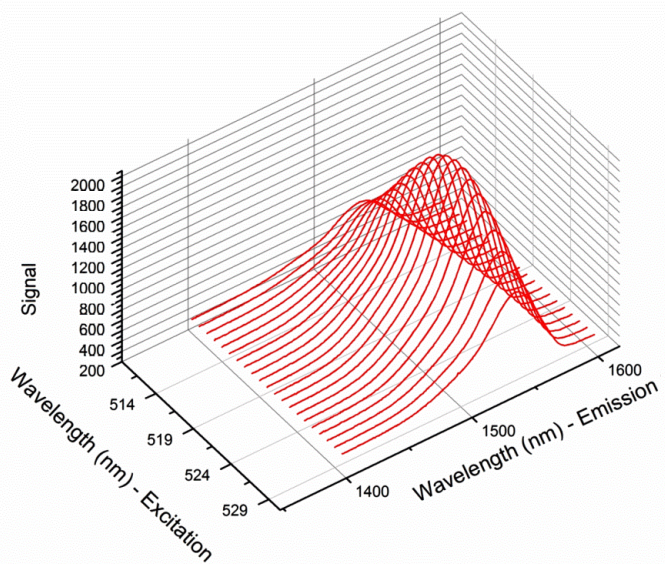

**Figure S7.** Photoemission excitation-emission matrix for $SiO_2$:Er film.

**Other Er-doped glasses**

Refractive index data have been reported for Er doping of glass materials prepared with various formulations and methods and discussed in the article [S9–S15]. The following are analyses of these data according to equation (1) of the article, treating $n_1$ as the undoped glass refractive index and $n_2$ as the Er-O inclusion refractive index. Values of $n_1$ and $n_2$ are obtained below by fitting data for $n$ vs. $f_2$.

$(Na_2O)_{0.25} (CaO)_{0.10} (SiO_2)_{0.65-x} (Er_2O_3)_x$ glass was prepared in [S9] by the melt-quench technique with x = 0 – 0.005. Data for the refractive index at 587.56 nm are shown figure S7 as blue circles. The solid curve models the Er-O inclusion as $f_2 = x/2$, yielding fitted parameters $n_1 = 1.525 \pm 0.003$ and $n_2 = 2.2 \pm 0.1$, and rms deviation 0.0005 between fit and data.

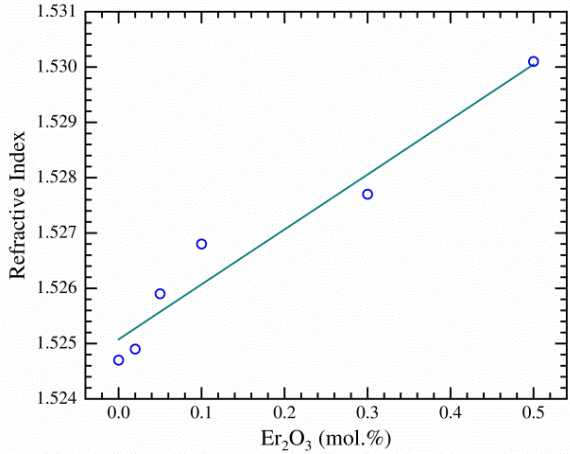 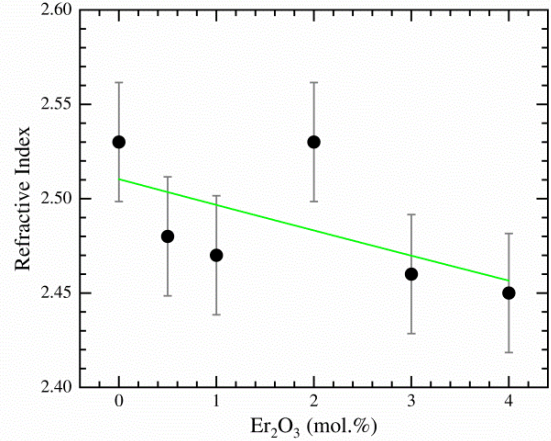

**Figure S8.** Refractive index of Na-Ca-Si-Er-O glass [S9]    **Figure S9.** Refractive index of Zn-B-Si-Er-O glass [S10]

$(ZnO)_{0.5-x}(B_2O_3)_{0.2}(SiO_2)_{0.3}(Er_2O_3)_x$ glass was prepared in [S10] by the melt-quench technique with x = 0 – 0.04. Date for the refractive index at unspecified wavelength is shown in figure S8 as black circles. The solid curve models the Er-O inclusion as $f_2 = x/2$, yielding fitted parameters $n_1 = 2.51 \pm 0.02$, $n_2 = 2.0 \pm 0.3$ and rms deviation 0.032 between fit and data, illustrated by error bars.

$GeO_2$ and $(GeO_2)_{0.8}(SiO_2)_{0.2}$ glass films with 0.2% $Er^{3+}$ doping were prepared in [S11] by a sol-gel method and annealed at 600 °C. Refractive indices measured at 633 nm are 1.62 and 1.58, respectively. Calculated refractive indices with the Er doping are 1.6060 and 1.5751, respectively, and without Er doping are 1.6054 and 1.5744, respectively, indicating changes with Er doping of 0.0006 and 0.0007, respectively, below the resolution of two decimal places in the presented data.

$(SiO_2)_{0.7}(HfO_2)_{0.3}$ films with $Er^{3+}$ dopings of 0.3, 0.5, and 1.0 mol. % were prepared in [S12] by sol-gel dip coating and annealing at 900 °C in air. Results show no changes in refractive indices to 3rd decimal place resolution at wavelengths of 632.9 and 543.5 nm, finding refractive indices for TE modes of 1.627 and 1.633, respectively. An increase in the refractive index by 0.003 for 1 mol. % Er doping is calculated from the effective medium model with $n_1 = 1.63$ and $n_2 = 1.95$.

$(SiO_2)_{0.8}(TiO_2)_{0.2}$ films with 0 and 0.5 % $ErO_{1.5}$ prepared in [S13] by sol-gel have refractive indices of 1.60 and 1.61, respectively, at full densification. The calculated increase in refractive index with Er doping, assuming $n_2 = 1.95$, is 0.0015 and therefore below experimental resolution ($\pm 0.01$ error is given).

$(GeO_2)_{0.8}(SiO_2)_{0.2}$ doped with 1 mol. % $Er^{3+}$ was prepared by spin-coating three layers of sol-gel material and annealing at 700 °C in [S14]. The refractive index is plotted in [S14] at six wavelengths with $\pm 0.005$ errors bars and the value of 1.489 is compared as similar (larger by about 0.009) to that of the undoped bulk glass. The erbium doping is expected to increase the refractive index by about 0.0019.

${[(TeO_2)_{0.7}(B_2O_3)_{0.3}]_{1-x}(ZnO)_x}_{1-y}(Er_2O_3)_y$ glasses were prepared in [S15] by the melt-quench technique with y = 0 – 0.05 mol. %. The authors attribute the increase in refractive index (at 632.80 nm), e.g. 1.7000 at y = 0 and 1.7400 at 0.05 mol. %, to the formation of non-bridging oxygens. Predicted change from the polarizability of $Er^{3+}$ is expected to be less than 0.0001.less than 0.0001.